\documentclass[aip,jap,amsmath,amssymb,reprint]{revtex4-1}

\usepackage{graphicx}
\usepackage{dcolumn}
\usepackage{bm}

\begin{document}
\title{Spin transport in antiferromagnetic NiO and magnetoresistance in Y$_3$Fe$_5$O$_{12}$/NiO/Pt structures} 
\author{Yu-Ming Hung}
\affiliation{Department of Physics, New York University, New York, New York 10003}
\author{Christian Hahn}
\affiliation{Department of Physics, New York University, New York, New York 10003}
\author{Houchen Chang}
\author{Mingzhong Wu}
\affiliation{Department of Physics, Colorado State University, Fort Collins, Colorado 80523}
\author{Hendrik Ohldag}
\affiliation{Stanford Synchrotron Radiation Lightsource, Menlo Park, California 94025}
\author{Andrew D. Kent}
\affiliation{Department of Physics, New York University, New York, New York 10003}
\date{\today}
\begin{abstract}
We have studied spin transport and magnetoresistance in yttrium iron garnet (YIG)/NiO/Pt trilayers with varied NiO thickness. 
To characterize the spin transport through NiO we excite ferromagnetic resonance in YIG with a microwave frequency magnetic field and detect the voltage associated with the inverse spin-Hall effect (ISHE) in the Pt layer. 
The ISHE signal is found to decay exponentially with the NiO thickness with a characteristic decay length of 3.9 nm. 
This is contrasted with the magnetoresistance in these same structures. 
The symmetry of the magnetoresistive response is consistent with spin-Hall magnetoresistance (SMR). 
However, in contrast to the ISHE response, as the NiO thickness increases the SMR signal goes towards zero abruptly at a NiO thickness of $\simeq 4$ nm, highlighting the different length scales associated with the spin-transport in NiO and SMR in such trilayers.
\end{abstract}
\maketitle 
\section{Introduction}
Antiferromagnets have attracted a great deal of attention in spintronics because of their unique properties, such as a low magnetic susceptibility and terahertz spin dynamics. \cite{Jungwirth_review,Baltz_review}
In metallic antiferromagnet-based spintronics, electrical switching of the antiferromagnetic domains in CuMnAs was sucesfully performed \cite{Science} and spin pumping studies have demonstrated spin injection into IrMn. \cite{IrMn, WeiZhang}
In oxide spintronics, ferromagnet (FM)/antiferromagnet (AFM)/heavy metal trilayer structures have been used to study spin-transport in insulating antiferromagnets. \cite{Hahn_SMR, Wang_PRL,Saitoh_CoO,Moriyama,Kai,Cheng,Cheng2,Hals} Experimental techniques include microwave-field-induced magnetization precession in the FM to generate a spin-current in the AFM and the inverse spin Hall effect (ISHE) in the heavy metal to convert the spin-current transmitted through the AFM into a voltage. \cite{Hahn_SMR, Wang_PRL,Saitoh_CoO,SMR_NiO} Such experimental studies have shown that NiO can be an efficient spin-conductor. A theoretical model has been proposed that explains these experimental results by spin-currents conducted by evanescent spin-waves in NiO. \cite{Slavin,Rezende} In similar structures, YIG/NiO/Pt, the spin-Hall magnetoresistance (SMR) has been measured as a function of temperature. Results suggest that NiO can suppress the magnetic proximity effect measured in a 3 nm thick Pt layer and were further interpreted to indicate spin-transport between Pt and YIG through NiO layers. \cite{SMR_NiO}

These two types of experiments have not been conducted on the same samples as a function of the NiO thickness. Here, we investigate spin transport through NiO layers with varied thickness at room temperature by exciting ferromagnetic resonance (FMR) in YIG and detecting the voltage signal across the Pt film associated with the ISHE. The SMR as a function of the NiO thickness was measured in the same samples. A comparison of these results shows that the ISHE signal decreases monotonically as a function of NiO thickness, while the SMR vanishes abruptly at a thickness of about 4 nm.
\section{Experimental methods}
YIG(20 nm)/NiO/Pt(5 nm) trilayers were deposited on gadolinium gallium garnet (Gd$_3$Ga$_5$O$_{12}$, GGG) substrates. YIG with a $\langle$111$\rangle$ orientation was deposited by magnetron sputtering. \cite{Houchen_YIG,Houchen_YIG2} The YIG film was subsequently exposed to ambient conditions. Ar$^+$ ion cleaning with a discharge voltage of 800 V was used prior to depositing NiO by radio frequency (rf) magnetron sputtering. A 5 nm thick Pt layer was deposited in-situ by electron beam evaporation. Samples with 0 (i.e. without NiO), 2, 4, 6, and 10 nm thick NiO layers were prepared and studied.

\begin{figure}[!t]
\centering
\includegraphics[width=3.4in]{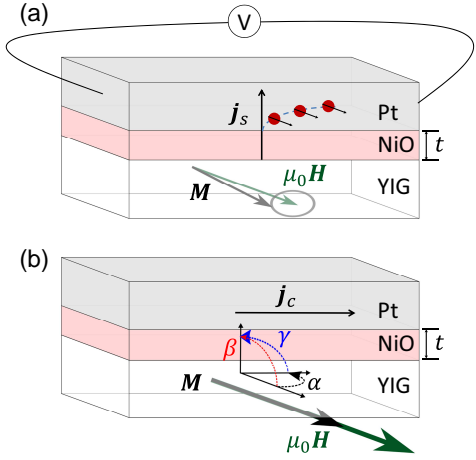}
\caption{Schematics of (a) inverse spin Hall voltage measurement by microwave-field-induced magnetization precession in YIG(20 nm)/NiO(t nm)/Pt(5 nm) and (b) Pt resistance measurement as a function of angle of applied field ($\alpha$, $\beta$, and $\gamma$) with respect to three different axes of the same sample.}
\label{fig1}
\end{figure}

For FMR studies, we mount the sample on a Cu stripline with a width of 500 $\mu$m, similar to the experimental geometry in Ref.~\citenum{Hahn_PRB}. We apply a microwave signal at 3.85 GHz and measure the inverse spin Hall voltage $V_{\textrm{ISHE}}$ across the Pt film as a function of the applied field using a lock-in technique. The microwave signal is turned on and off at a few kHz and the lock-in response at this frequency is measured. Figure \ref{fig1}(a) shows a schematic of the sample and applied field geometry, while Fig. \ref{fig1}(b) shows a schematic of the magnetoresistance experiments. Magnetoresistance measurements of YIG/NiO/Pt trilayers were performed in a 4-wire configuration, i.e. with separate voltage and current contacts in a line. We apply an external magnetic field of 0.2 T to align the magnetization of YIG with the field. The samples are then rotated around three different axes to obtain the magnetoresistance as a function of angle. All the experiments presented in this paper were conducted at room temperature.
\section{Results}
\begin{figure}[!t]
\centering
\includegraphics[width=3.4in]{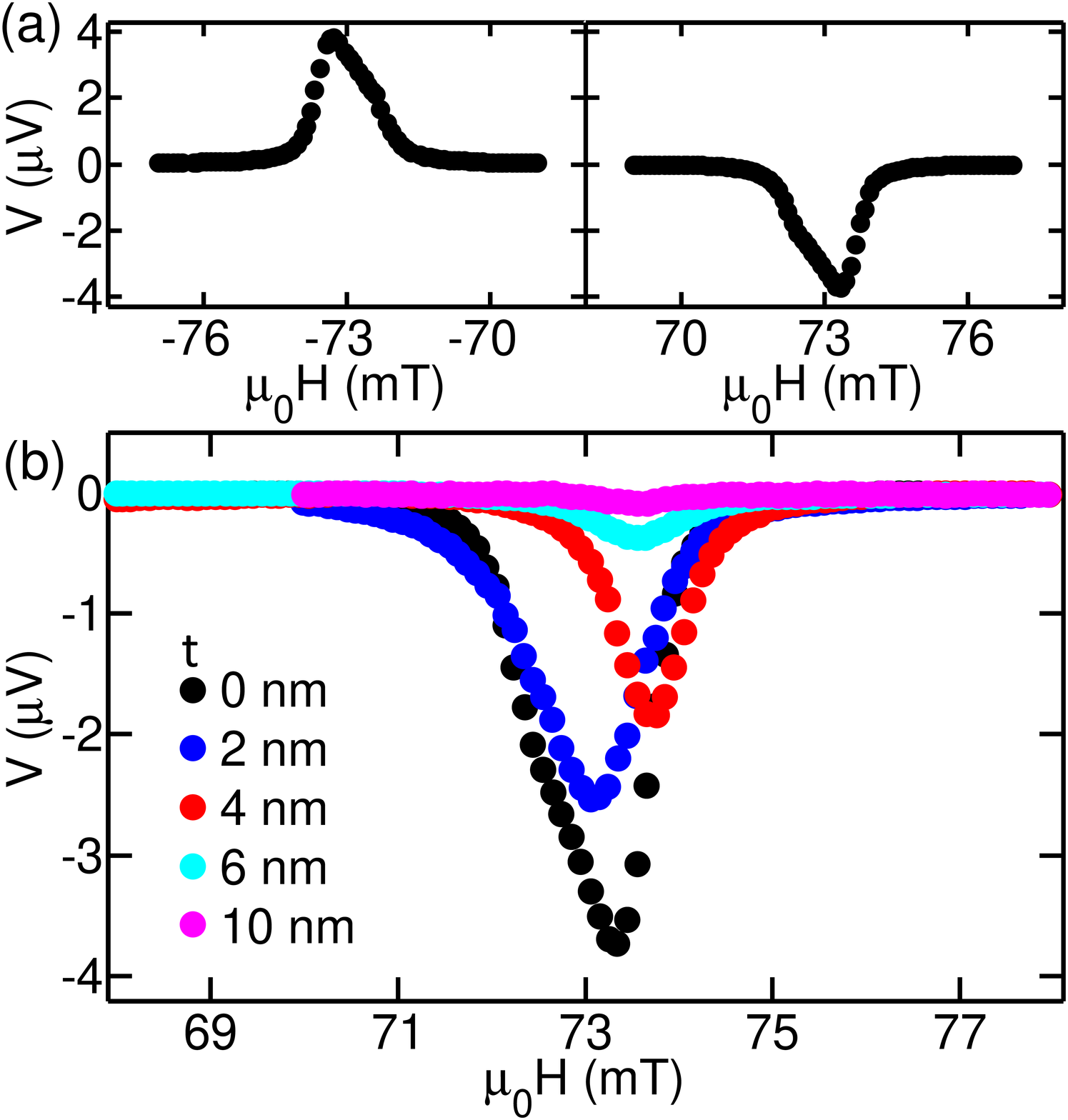}
\caption{ Inverse spin Hall voltage $V_{\textrm{ISHE}}$ measured as a function of external field in (a) YIG(20 nm)/Pt(5 nm) and (b) YIG(20 nm)/NiO(t nm)/Pt(5 nm) with a 3.85 GHz rf excitation.}
\label{fig2}
\end{figure}

First we present measurements of the ISHE voltage in YIG(20 nm)/Pt(5 nm), thin films that do not have a NiO layer. Fig. \ref{fig2}(a) shows the ISHE voltage as a function of the applied field. A peak in the ISHE voltage signal is seen at the resonance field of the YIG. (The FMR absorption data is not shown.) There is also a minor peak which is most likely due to inhomogneity in our extend film samples. As previously reported in Ref.~\citenum{Hahn_PRB,Mosendz,Kajiwara}, the measured ISHE voltage $V_{\textrm{ISHE}}$ is an odd function of the magnetic field which is a signature that the effect is based on spin pumping into Pt and not a thermoelectric signal. In YIG/NiO/Pt trilayers the $V_{\textrm{ISHE}}$ as a function of NiO thickness is shown in Fig. \ref{fig2}(b). We find slightly different resonance fields on different YIG samples, which may be associated with small variation in their magnetization. We checked that the shift in resonance field is symmetric in both field directions indicating that it is not caused by exchange bias introduced in the YIG by NiO. We extract the peak values of $V_{\textrm{ISHE}}$ and plot it as a function of NiO thickness in Fig. \ref{fig4}. The $V_{\textrm{ISHE}}$ signal in YIG/NiO/Pt indicates that exciting FMR in YIG is able to produce spin-currents in NiO and spin-injection into Pt.

\begin{figure}[!t]
\centering
\includegraphics[width=3.4in]{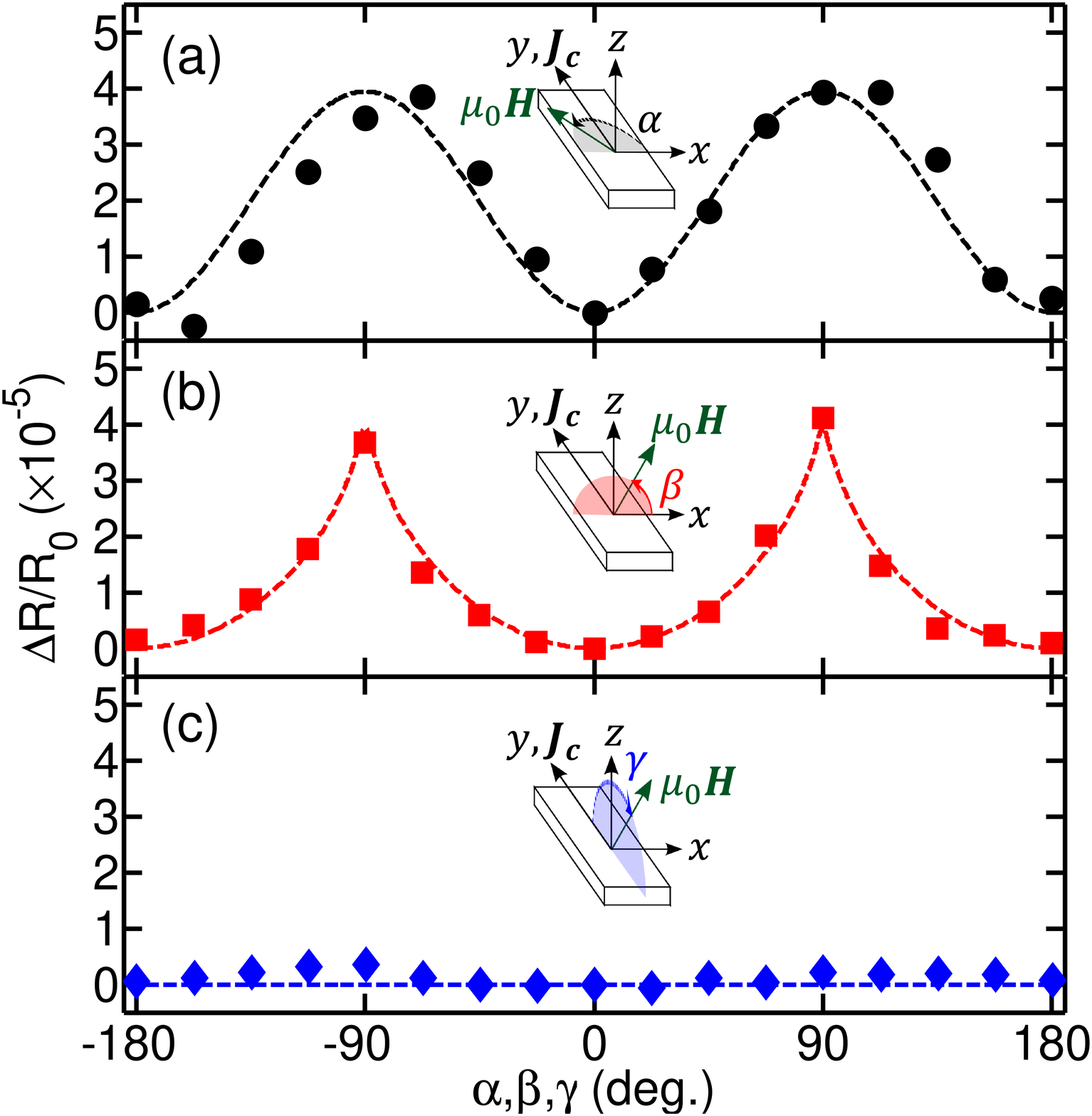}
\caption{Magnetoresistance of YIG(20 nm)/NiO(2 nm)/Pt(5 nm) as a function of the field angle rotated in the (a) x-y plane an angle $\alpha$, (b) x-z plane an angle $\beta$, and (c) y-z plane an angle $\gamma$. The dashed line is the expected SMR response based on a macrospin model.}
\label{fig3}
\end{figure}

\begin{figure}[!t]
\centering
\includegraphics[width=3.4in]{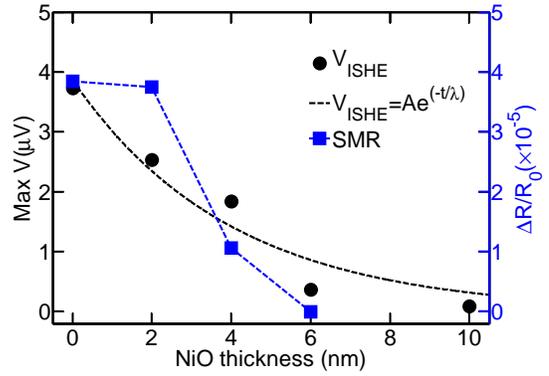}
\caption{Black circles are the peak values of the ISHE voltage extracted from Fig. \ref{fig2}(b), while blue squares are the maximum values of SMR $\Delta R_{\textrm{Max}}/R_0$ extracted from Fig. \ref{fig3}(b) for different NiO thicknesses. Black dashed line is a fit to A$e^{-t/ \lambda}$. Blue dashed line is only a guide to the eyes.}
\label{fig4}
\end{figure}

We now discuss the magnetoresistance measurements of YIG/Pt and YIG/NiO/Pt. Magnetoresistance as a function of angle of applied field with respect to three different axes of the sample were measured. In these experiments, the applied field $\mu_0H$ = 0.2 T is fixed. Figure \ref{fig3} shows the magnetoresistance of YIG(20 nm)/NiO(2 nm)/Pt(5 nm) as a function of angle $\alpha$ (rotation in the x-y plane) in Fig. \ref{fig3}(a), $\beta$ (rotation in the x-z plane) in Fig. \ref{fig3}(b), and $\gamma$ (rotation in the y-z plane) in Fig. \ref{fig3}(c). The insets in Fig. \ref{fig3} show the experimental geometry and definitions of angles, $\alpha$, $\beta$, and $\gamma$. 

We measured the magnetoresistance with a resolution of $\sim 10^{-6}$ in all three geometries. If we assume SMR and anisotropic magnetoresistance (AMR) as possible contributions to the measured magnetoresistance, the angle dependent measurements in Fig. \ref{fig3}(a), \ref{fig3}(b), and \ref{fig3}(c) should represent contributions of SMR and AMR, SMR only, and AMR only, respectively. The signal observed in Fig. \ref{fig3}(c) is of the order of the measurement noise. Given the measurement noise is  $\sim 10^{-6}$, the peak value of the signal is at least a factor of 10 smaller than the SMR signal. This is consistent with the results reported in Ref.~\citenum{SMR_NiO}, where it was concluded based on no AMR response that there are no induced magnetic moments in Pt with a NiO interlayer. However, we also find negligible signal when rotating in the y-z plane to sweep the angle $\gamma$ for the YIG(20 nm)/Pt(5 nm) sample, which does not have the NiO interlayer. The AMR signal indicative of magnetic proximity effect found in Ref.~\citenum{SMR_NiO} was measured in a sample with 3 nm Pt thickness as opposed to the thicker films used here and in Refs.~\citenum{Hahn_PRB}, \!\!\citenum{SMR_Saitoh}, where also no AMR was reported. If magnetic proximity effect is present in our samples, its contribution to the magnetoresistance is negligible compared to the contribution of SMR. Therefore, the resistance variations in Figs. 3(a) and 3(b) can be attributed to the SMR, which describes how the Pt resistance reflects the itinerant electron-spin interactions at the NiO/Pt interface. The similar magnitude ($\Delta R_{\textrm{Max}}/R_0 \simeq 4\times10^{-5}$) in Figs. 3(a) and 3(b) also suggest we only have a SMR signal. We observe a periodic response with a period of 180 degrees in Figs. 3(a)(b) which can be described with $R-R_0=\Delta R=\Delta R_{\textrm{Max}}\sin ^2\theta$. $\theta$ is the angle between the magnetization $\textbf{M}$ in YIG and the spin polarization from the spin Hall effect in Pt. We compute the equilibrium magnetization direction as a function of the applied field direction, $\alpha$ and $\beta$, based on a macrospin model considering the demagnetization field and external field. The results are $\alpha = \theta$ in Fig. \ref{fig3}(a) and $\beta = \theta+\arcsin(\sin(2\theta)M_s/2H)$ in Fig. \ref{fig3}(b). The sharp peaks in the SMR signal at = +/-90 degree seen in Fig. \ref{fig3}(b) are due to the difference between the angles of the applied field ($\beta$) and the YIG magnetization ($\theta$). The magnetization of YIG was determined by ferromagnetic resonance characterization to be  $\mu_0M_s=0.176$ T. As seen in the form of dashed lines in Fig. \ref{fig3}, plots of $\Delta R/R_0$ versus field angle ($\alpha$, $\beta$, and $\gamma$) fit the experimental data well.

The results in Fig. \ref{fig3} show that the direction of YIG magnetization can affect the resistance measured in Pt. This suggests that the YIG magnetization rotates the spins in the NiO which changes the spin-scattering and accumulation at the NiO/Pt interface and thus the resistance of the Pt film. We repeat the same experiment for YIG/Pt and YIG/NiO/Pt with t = 4 and 6 nm to extract the maximum value $\Delta R_{\textrm{Max}}/R_0$ in Fig. \ref{fig3}(b) and plot it as a function of NiO thickness in Fig. \ref{fig4}.

We now compare the two experiments as a function of NiO thickness. We show the ISHE voltage of YIG/NiO/Pt as a function of NiO thickness and plot it using black circles on the left hand axis in Fig. \ref{fig4}. We fit the $V_{\textrm{ISHE}}$ data points with $Ae^{-t/\lambda}$ and obtain a decay length of 3.9 nm. The blue squares plotted with reference to the right axis in Fig. \ref{fig4} represent the SMR signal of YIG/NiO/Pt as a function of NiO thickness. We see from the comparison in Fig. \ref{fig4} how both the ISHE voltage and SMR decrease monotonically with NiO thickness. In contrast to the ISHE signal, the SMR shows an abrupt decrease at a thickness of 4 nm. This points to a difference in how the critical length scales are determined in the two effects. In both cases spin flip-scattering of electrons in Pt on the NiO interface is involved. However, the exponential decrease associated with diffusion of magnons in NiO \cite{Slavin,Rezende} is not reproduced in the SMR data. On the contrary the insertion of a 2 nm NiO layer barely reduces the SMR amplitude. We can thus speculate that the source of the SMR, collinearity or non-collinearity between magnetic moments at the Pt/NiO interface and the electron spins, is abrubtly decoupled from the YIG magnetization above a certain NiO thickness.
\section{Conclusion}
In conclusion, we have reported ISHE voltage and SMR measurements in YIG/NiO/Pt layered structures. From the dependence on the NiO thickness, we found a monotonic decrease of the $V_{\textrm{ISHE}}$ and SMR signal with the NiO thickness. The sharp decrease of the SMR signal at t = 4 nm and the exponential decrease of the $V_{\textrm{ISHE}}$ signal suggests that the length scales of the phenomena are distinct. The abrupt decrease of the SMR signal may reflect the blocking of the NiO moments but this remains to be tested.
\begin{acknowledgments}
The growth of the YIG films at Colorado State University was supported by the SHINES, an Energy Frontier Research Center funded by the U.S. Department of Energy, Office of Science, Basic Energy Sciences under Award SC0012670. The research at NYU was sponsored by the Institute for Nanoelectronics Discovery and Exploration (INDEX), a funded center of Nanoelectronics Research Initiative (NRI), a Semiconductor Research Corporation (SRC) program sponsored by NERC and NIST.
\end{acknowledgments}
\bibliography{Bib_NiO}
\end{document}